# Compact mode-locked Er-doped fiber laser for broadband cavity-enhanced spectroscopy

Aleksander Głuszek[1], Francisco Senna Vieira[2], Arkadiusz Hudzikowski[1], Adam Wąż[1], Jarosław Sotor[1], Aleksandra Foltynowicz[2], and Grzegorz Soboń[1,*]

[1]*Laser & Fiber Electronics Group, Faculty of Electronics, Wrocław University of Science and Technology, Wybrzeze Wyspianskiego 27, 50-370 Wrocław, Poland*
[2]*Department of Physics, Umeå University, 901 87 Umeå, Sweden*
*Corresponding authors: grzegorz.sobon@pwr.edu.pl, aleksandra.foltynowicz@umu.se*

**Abstract:**

We report the design and characteristics of a simple and compact mode-locked Er-doped fiber laser and its application to broadband cavity-enhanced spectroscopy. The graphene mode-locked polarization maintaining oscillator is thermally stabilized and equipped with three actuators that control the repetition rate: fast and slow fiber stretchers, and metal-coated fiber section. This allows wide tuning of the repetition rate and its stabilization to an external reference source. The applicability of the laser to molecular spectroscopy is demonstrated by detecting $CO_2$ in air using continuous-filtering Vernier spectroscopy.

## 1. Introduction

Optical frequency combs emerged as powerful tools for molecular spectroscopy due to their high spectral brightness, broad spectral coverage, and compatibility with enhancement cavities [1]. In trace gas sensing, frequency combs allow detection of multiple species with high sensitivity and selectivity in short acquisition times [2-4]. Therefore there is an ongoing effort to harness the potential of comb spectroscopy for field-deployable gas detection platforms. Operation outside the laboratory requires compact, low-cost and low power consumption comb sources, which are enabled by the use of fiber-based technology [5,6]. Recently a fully self-referenced frequency comb based on an Er-doped fiber laser was reported, with electrical power consumption of only 5 watts [7]. It has been also proven that fiber-based frequency combs are ready for space travel and resistant to microgravity conditions [8]. Moreover, near-infrared fiber lasers can be used as pumps for compact mid-infrared combs based on nonlinear conversion [3,9].

Near and mid-infrared dual-comb spectrometers have been used for open path gas sensing using laboratory based platforms [10-12] and a dual-comb spectrometer based on Er-doped fiber lasers has been deployed in the field for open path monitoring of emissions [13] over large areas. However, a field-deployed platform for in-situ sensing has only been demonstrated in the UV range using a system based on a frequency-doubled Ti:Sapphire laser, an enhancement cavity, and low-resolution compact spectrograph [14,15]. Similar demonstrations in the near- and mid-infrared range are missing. In situ detection requires path enhancement in the form of a multipass cell or enhancement cavity. Cavity-enhanced dual comb spectroscopy [16] requires either complex stabilization schemes [17] or adaptive sampling [18]. A simpler alternative is continuous-filtering Vernier spectroscopy (CF-VS) [19,20], a cavity-enhanced technique in which the entire spectral bandwidth of the comb is recorded in tens of ms. The introduction of an offset between the repetition rate of the laser ($f_{rep}$) and the external cavity free spectral range (FSR) generates a series of transmission peaks within the comb spectrum – the so-called Vernier orders (VO). The width of each VO, which determines the spectral resolution, depends on the $f_{rep}$/FSR mismatch. The VOs are swept across the comb spectrum by scanning either the comb $f_{rep}$ or the cavity FSR, and a selected VO is recorded in cavity transmission. Scanning the $f_{rep}$ has the advantage that it allows using monolithic fixed-length cavities, with reduced risk of misalignment during operation. Slow and wide range $f_{rep}$ control is needed to introduce the initial $f_{rep}$/FSR offset (usually of the order of a few kHz), and faster sweep of $f_{rep}$ (at a rate of few tens of Hz with an amplitude of a few hundred Hz) is needed for spectral acquisition. The offset frequency of the laser does not need to be controlled. Therefore, the Vernier spectrometer can be used in combination with compact mode-locked fiber lasers.

Here we report a simple and compact 125 MHz mode-locked Er-doped fiber laser with wide $f_{rep}$ tuning capabilities and its application to broadband cavity-enhanced continuous-filtering Vernier spectroscopy. The laser is equipped with three different actuators for tuning of the $f_{rep}$, based on PZT fiber stretchers and a metal-coated fiber section acting as a resistive heater. The PZTs enable $f_{rep}$ tuning by 2.6 kHz in total and locking it to an external reference with relative rms of the noise of 0.42 mHz over >20 hours of measurement. The applicability of the laser to optical frequency comb spectroscopy was verified by detection of $CO_2$ in laboratory air using a CF-VS spectrometer, with relative precision of concentration determination similar to that obtained with a commercial Er-doped fiber frequency comb.

## 2. Laser setup and characteristics

The setup of the compact Er-doped fiber laser is depicted in Fig. 1a. The laser is built entirely of polarization maintaining (PM) fibers and components. The resonator consists of a filter-type wavelength division multiplexer (PM WDM), a hybrid component comprising an output coupler and isolator (OC/Isolator), a graphene-based saturable absorber, two PZTs and a piece of metal-coated optical fiber. The oscillator is pumped by a 980 nm single-mode laser diode. For efficient thermal stabilization and tuning, the laser is placed on a metal core printed circuit board (MCPCB) heating plate with uniformly spread heating traces, ensuring perfectly homogenous heat distribution. The whole system (the oscillator with thermal controller) consumes less than 5 W of power during operation. The resonator together with the MCPCB plate are placed in a 3D-printed enclosure (see Fig. 1b). Two piezo stretchers are used for $f_{rep}$ tuning: a multilayer piezo stack (5x5x50 mm) with 2.7 µF capacity and 40 µm maximum nominal displacement (labeled as "Slow PZT" in Fig. 1), and a single piezo chip with 160 nF capacity and 1.8 nm nominal displacement (labeled as "Fast PZT" in Fig. 1). Additional tuning is provided by resistive heating of a 60 mm-long segment of fiber coated with a thin layer of metal alloy (Ti/Ni/Au with 36 Ω resistance and maximum applicable voltage of 2.5V).

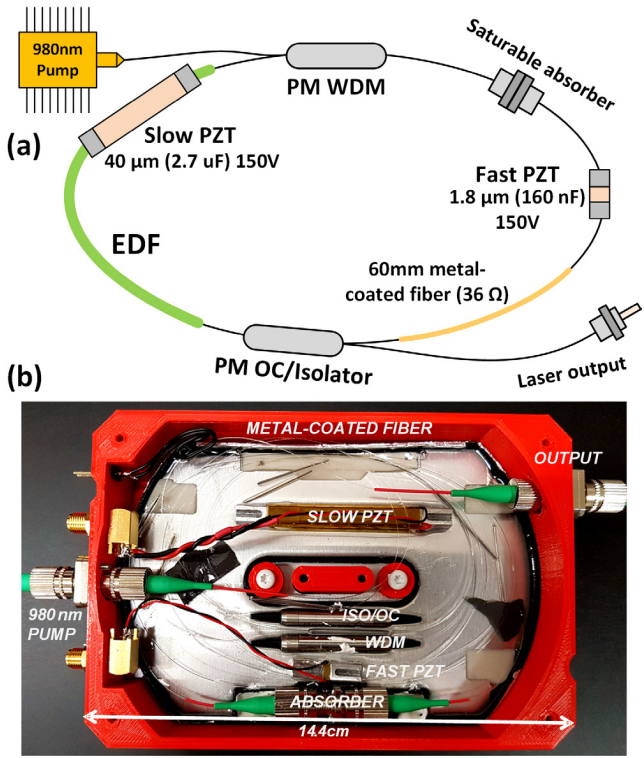

Fig. 1. (a) Experimental setup of the compact laser oscillator. PM WDM: polarization maintaining wavelength division multiplexer; EDF: erbium-doped fiber; PM OC/ISO: output coupler with isolator; PZT: piezoelectric transducer. (b) Photograph of the laser in the 3D-printed enclosure.

The characteristics of the laser are summarized in Fig. 2. Due to the net anomalous dispersion of the resonator, the laser generates a soliton-shaped spectrum with 8.08 nm full width at half maximum (FWHM) bandwidth centered at 1564 nm (Fig. 2a). The pulse duration measured with an intensity autocorrelator is 335 fs assuming a sech² pulse shape (Fig. 2b). Figure 2c shows the radio-frequency (RF) spectrum in the full available span of the spectrum analyzer (Keysight N9010A, 3.6 GHz), indicating stable mode-locking without any parasitic modulations. The pulse repetition rate without any voltage applied to the PZT actuators and the temperature set to 30°C is equal to 125.028 MHz. The first harmonic of the RF spectrum is shown in Fig. 2d, with signal to noise ratio of more than 70 dB.

Figure 3 shows the static tuning performance of the laser repetition rate using the three available actuators and temperature control. The RF spectra show the first harmonic (~125.028 MHz) recorded for two extreme setpoints of each actuator. The slow PZT provides elongation of the fiber by ~37 µm under maximum driving voltage (150 V) which corresponds to a repetition rate change of 2.5 kHz (Fig. 3a) and 3.8 GHz comb mode tuning in the optical domain. The fast PZT elongates the fiber by ~1.5 µm, resulting in $f_{rep}$ change by 103 Hz (Fig. 3b) and comb mode shift by 158 MHz in the optical domain. The difference between the nominal PZT displacement and the obtained optical path elongation results from the gluing process, where bond elasticity and piezo pre-tension play the main role. A relatively large tuning range of 536 Hz (corresponding to 8 µm of static stretch) is provided by the metal-coated resistive heater (Fig. 3c) by applying only 2.5 V. The $f_{rep}$ can be widely tuned by heating the MCPCB plate: by changing the temperature of the laser from 30°C to 50°C the repetition rate decreases by 17 kHz, corresponding to a thermal tuning coefficient of -0.85 kHz/°C.

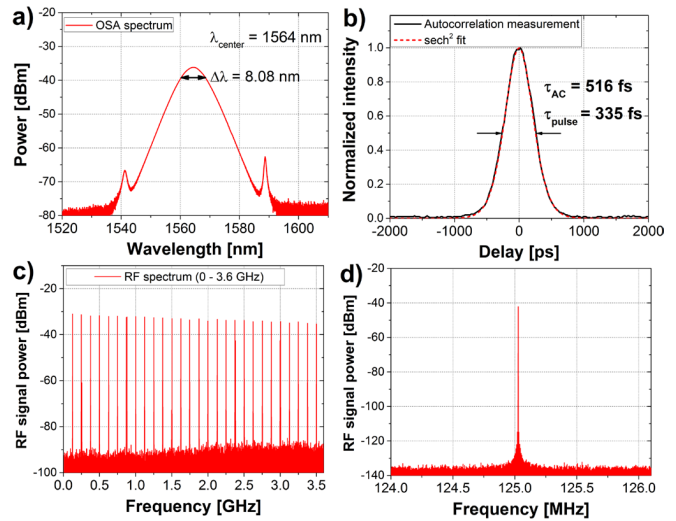

Fig. 2. Characteristics of the oscillator: (a) optical spectrum measured with 0.01 nm resolution, (b) pulse autocorrelation indicating a duration of 335 fs, (c) radio-frequency spectrum measured with 270 kHz resolution bandwidth and 3.6 GHz frequency span, (d) narrow-span RF spectrum showing the first harmonic, measured with 10 Hz resolution bandwidth.

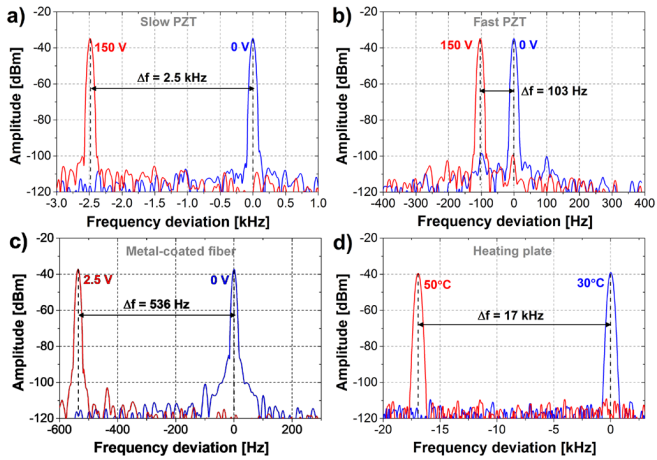

Fig. 3. Static tuning characteristics of the laser repetition rate using various actuators: (a) slow PZT, (b) fast PZT, (c) metal-coated fiber, (d) MCPCB heating plate. The zero frequency corresponds to the nominal repetition frequency with no voltage applied to any actuator and the temperature set to 30°C (125.028 MHz).

The frequency response of the actuators was experimentally verified using a fiber-based laser vibrometer, as schematically depicted in Fig. 4a. The vibrometer operates at the wavelength of 1550 nm [21] and is capable of measuring displacements with resolution down to 1 nm in the 0 – 1 MHz frequency range. Figure 4b shows the obtained mechanical stretch of the investigated actuators (both PZTs and the metal-coated fiber) as a function of the modulation frequency, expressed in meters per 1 V of applied voltage. During the measurement, the PZTs were driven with 3.56 $V_{rms}$ sine wave from a signal generator while the metal-coated fiber was driven with 1.4 $V_{rms}$. The measurement revealed a resonance for both PZTs at 8 kHz and a flat response between 0 to 5 kHz. The response of the metal-coated fiber heater decreases with increasing frequency, however, the maximum possible thermally induced stretching of the 60 mm coated segment of fiber is at the level of 8 μm for static conditions.

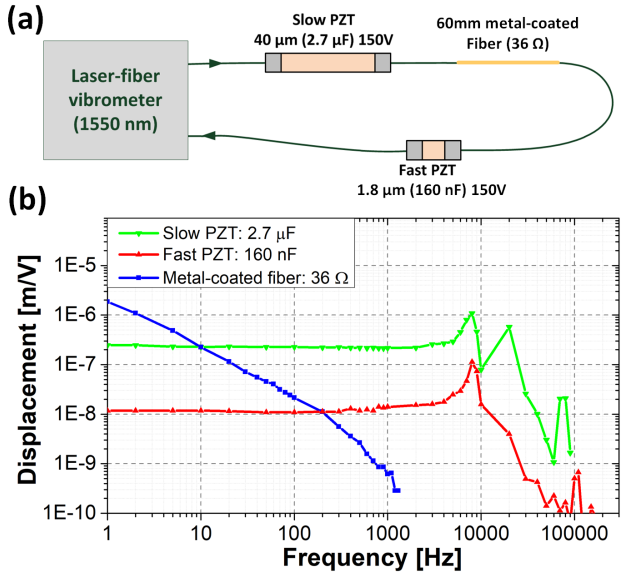

Fig. 4. Actuator bandwidth measurement using laser-fiber vibrometer: (a) experimental setup, (b) response of the investigated actuators per one volt as a function of the modulation frequency.

The stability of the oscillator was investigated by measuring the repetition frequency over a period of 22 hours in two cases: without active $f_{rep}$ stabilization (with only thermal stabilization of the MCPCB at 32 °C) and with $f_{rep}$ locked to an external reference. The passive long term stability measurement is plotted in Fig. 5a. The $f_{rep}$ was measured by an RF signal analyzer (Keysight EXA N9010A) referenced to its internal clock source (±2ppm in the 20-30°C range). Data was acquired every 30 s with 10 Hz resolution bandwidth (RBW). It can be observed that the simple temperature stabilization based on the MCPCB board and a 3D-printed enclosure provides a very good stability of the $f_{rep}$ (amplitude of the drift does not exceed 150 Hz). A stability improvement can be observed between hours 12 – 18 of the measurement, which corresponds to night hours (2 am – 8 am) and a steady temperature in the laboratory. The frequency variation is caused mainly by the thermal instabilities of the oscillator (the temperature controller stability, temperature gradient in the laser enclosure, etc.) and the inaccuracy of the internal clock source in the RF analyzer. For active stabilization a signal generator (HP 8648C) was locked to the RF analyzer using the 10 MHz reference output. A 125.028 MHz signal was generated and compared with the oscillator $f_{rep}$ by a phase detector (Menlo DXD200). The error signal was delivered to a servo controller (Vescent Photonics D2-125) that drove the PZT stretchers via a 0-150 V amplifier (Piezomechanik SVR 150/3). For better resolution, the $f_{rep}$ was measured at the 25th harmonic (3.126 GHz) and normalized to the first harmonic (i.e. divided by 25). Data points were acquired every 30 s with 10 Hz span and 1 Hz RBW. Figure 5b presents the relative stability of the oscillator when the $f_{rep}$ was locked to the reference clock in the RF analyzer. The rms of the noise is <0.42 mHz demonstrating the excellent relative stability of the repetition frequency in long term operation.

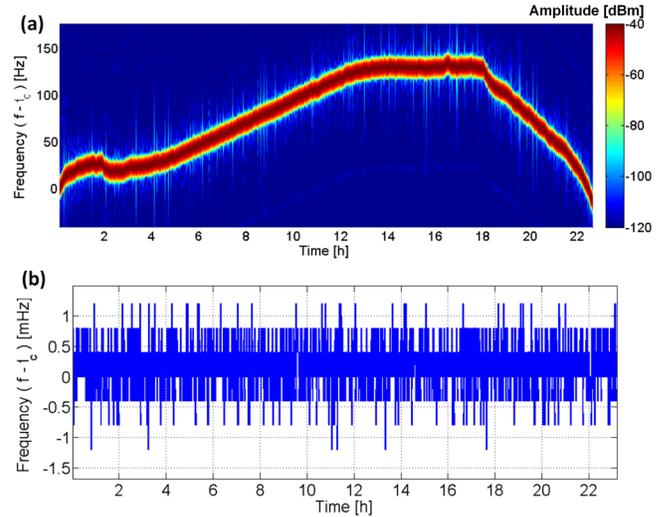

Fig. 5. Repetition frequency measured over 22 hours with: (a) a free-running laser (with only thermal stabilization of the cavity at 32°C), (b) laser stabilized to an RF reference. $f_c$ – center frequency of 125.028 MHz.

## 3. Cavity-enhanced Vernier spectroscopy

The applicability of the compact Er-doped fiber laser to broadband spectroscopy was verified by using it as a source for continuous-filtering Vernier spectrometer, depicted in Fig. 6. The light emitted by the oscillator was amplified in an all-fiber erbium-doped fiber amplifier (EDFA) to broaden the spectrum via self-phase modulation in order to cover the absorption band of $CO_2$. The output of the amplifier was sent through a polarization maintaining fiber isolator, coupled to free space and mode-matched to a 60-cm long linear cavity (FSR = 250 MHz, finesse 1050) by a telescope. The cavity was open to air; alternatively a 50-cm tube was placed inside the cavity and filled with $N_2$ to provide absorption-free background spectra. A relative mismatch of 6.7 kHz between $f_{rep}$ and FSR/2, corresponding to a spectral resolution of 4.4 GHz, was introduced either by detuning the cavity length by 32 μm from the perfect match condition (i.e. FSR = $2f_{rep}$) using a translation stage (TS) or by changing the laser heater temperature by 8 °C. The output beam of the cavity was incident on a diffraction grating (DG, 600 grooves/mm) mounted on a galvo-scanner (GS). A half-wave plate (λ/2) placed before the DG was used to adjust the polarization to maximize the diffraction efficiency. The selected VO was swept across the spectrum by applying a 20-Hz sinewave from a function generator (FG) to the laser slow PZT via a high voltage amplifier (HVA). To preserve the propagation direction of the selected VO during the sweep, the GS was simultaneously scanned at the same frequency, with the phase and amplitude of the sinewave adjusted for optimal synchronization. The remaining mismatch was actively corrected by stabilizing the beam direction after the diffraction grating via a feedback loop actuating on the laser $f_{rep}$ [19,22,23]. The diffracted beam was sent through an aperture to block the neighboring VOs, and afterwards split into three arms by two beam splitters (BS 1 & 2) for spectral acquisition (PD 1), frequency calibration ($CaF_2$ & PD 2), and stabilization of the spectral scan (QD). The horizontal difference signal of the quadrant detector (QD) was used as an error signal input to a proportional integral (PI) controller. The correction signal from the PI controller was summed with the sinewave that drove the laser PZT ensuring relative stabilization of the PZT and galvo scans. The outputs of photodetectors PD 1 & 2 were recorded by a multichannel data acquisition card (DAQ) with 100 kS/s acquisition rate. The relative frequency calibration was realized by resampling the spectrum at the zero crossings of the $CaF_2$ etalon (3-mm long, FSR = 35 GHz) fringes, detected by PD 2, as described in [23]. Absolute calibration was achieved by comparing the spectrum to a model calculated using a line list from the HITRAN 2016 database [24].

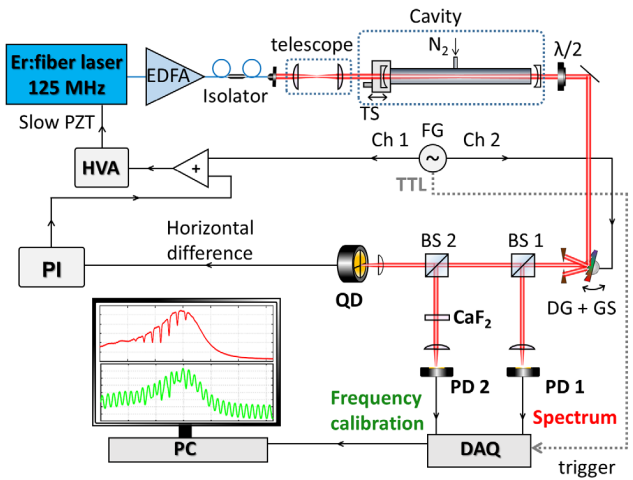

Fig. 6. Experimental setup of the continuous-filtering Vernier spectrometer based on the compact Er-doped fiber oscillator. EDFA: Erbium-doped fiber amplifier; HVA: high voltage amplifier; TS: translation stage; λ/2: Half-wave plate; DG: diffraction grating; GS: galvo scanner; BS 1 & 2: beam splitters; PD 1 & 2: InGaAs photodiodes; QD: quadrant detector; PI: proportional integral controller; FG: function generator; Ch 1 & 2: channels of the function generator; DAQ: multichannel data acquisition card.

Figure 7a shows a normalized absorption spectrum of $CO_2$ in air (markers) together with a fit of a CF-VS model [20,22] calculated using a line list from the HITRAN 2016 database, using only $CO_2$ concentration as a fitting parameter. The retrieved $CO_2$ concentration is 694(6) ppm, with the precision given by the standard deviation of 20 consecutive scans. Overall, the data agrees well with the model, and the structure visible in the residual of the fit (Fig. 7b) is mainly caused by inaccuracies in the frequency calibration procedure, as previously observed in [22]. Nevertheless, the precision in concentration determination is similar to that obtained in a previous work using CF-VS system based on a commercial Er-doped fiber frequency comb [23] indicating the potential of our compact source for quantitative measurements.

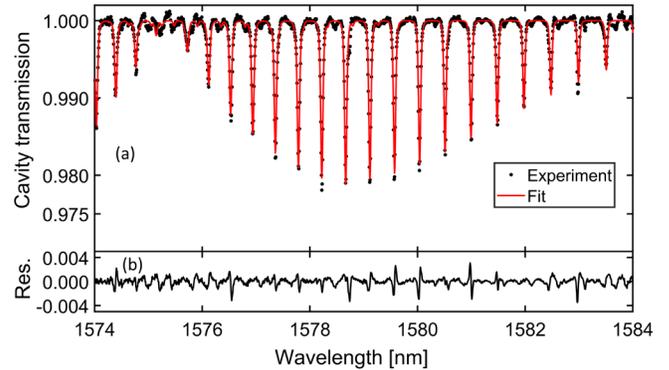

Fig. 7. (a) Experimental spectrum of the R branch of the $3v_1 + v_3$ $CO_2$ band in laboratory air (black markers) compared to a fit of a Vernier spectrum model (red line); (b) Absolute residuum of the fit

## 4. Summary

We demonstrated a compact and simple mode-locked Er-doped fiber oscillator with active $f_{rep}$ stabilization and temperature control that consumes less than 5 W of power. The repetition rate can be widely tuned using three actuators. The implemented PZT stretchers enable efficient locking of the repetition rate to an external reference source, with relative noise RMS value better than 0.42 mHz over 22 hours. The applicability of the laser to cavity-enhanced spectroscopy was verified by detection of $CO_2$ in atmospheric air using a continuous-filtering Vernier spectrometer. Ongoing improvements to the Vernier spectrometer that will remove the need for active stabilization of the scan will enable field deployment of the platform for in-situ atmospheric measurements.

## Acknowledgements

The work was funded by the Foundation for Polish Science (FNP, First TEAM/2017-4/39), the Polish National Agency For Academic Exchange NAWA (PPI/APM/2018/1/00031/U/001), and Knut and Alice Wallenberg Foundation (KAW 2015.0159).

## References

1. N. Picqué, T.W. Hänsch, "Frequency comb spectroscopy," Nature Photon. **13**, 146–157 (2019).
2. F. Adler, P. Masłowski, A. Foltynowicz, K. C. Cossel, T. C. Briles, I. Hartl, and J. Ye, "Mid-infrared Fourier transform spectroscopy with a broadband frequency comb," Opt. Express **18**, 21861-21872 (2010).
3. G. Ycas, F. R. Giorgetta, E. Baumann, I. Coddington, D. Herman, S. A. Diddams, and N. R. Newbury, "High-coherence mid-infrared dual-comb spectroscopy spanning 2.6 to 5.2 μm," Nature Photon. **12**, 202-208 (2018).

4. A. V. Muraviev, V. O. Smolski, Z. E. Loparo, and K. L. Vodopyanov, "Massively parallel sensing of trace molecules and their isotopologues with broadband subharmonic mid-infrared frequency combs," Nature Photon. **12**, 209-214 (2018).
5. L. C. Sinclair, J.-D. Deschênes, L. Sonderhouse, W. C. Swann, I. H. Khader, E. Baumann, N. R. Newbury, and I. Coddington, "Invited Article: A compact optically coherent fiber frequency comb," Rev. Sci. Instrum. **86**, 081301 (2015).
6. L. C. Sinclair, I. Coddington, W. C. Swann, G. B. Rieker, A. Hati, K. Iwakuni, and N. R. Newbury, "Operation of an optically coherent frequency comb outside the metrology lab," Opt. Express **22**, 6996-7006 (2014).
7. P. Manurkar, E. F. Perez, D. D. Hickstein, D. R. Carlson, J. Chiles, D. A. Westly, E. Baumann, S. A. Diddams, N. R. Newbury, K. Srinivasan, S. B. Papp, and I. Coddington, "Fully self-referenced frequency comb consuming 5 watts of electrical power," OSA Continuum **1**, 274-282 (2018).
8. M. Lezius, T. Wilken, C. Deutsch, M. Giunta, O. Mandel, A. Thaller, V. Schkolnik, M. Schiemangk, A. Dinkelaker, A. Kohfeldt, A. Wicht, M. Krutzik, A. Peters, O. Hellmig, H. Duncker, K. Sengstock, P. Windpassinger, K. Lampmann, T. Hülsing, T. W. Hänsch, and R. Holzwarth, "Space-borne frequency comb metrology," Optica **3**, 1381-1387 (2016).
9. K. Krzempek, D. Tomaszewska, A. Głuszek, T. Martynkien, P. Mergo, J. Sotor, A. Foltynowicz, and G. Soboń, "Stabilized all-fiber source for generation of tunable broadband $f_{CEO}$-free mid-IR frequency comb in the 7 – 9 μm range," Opt. Express **27**, 37435-37445 (2019).
10. G. B. Rieker, F. R. Giorgetta, W. C. Swann, J. Kofler, A. M. Zolot, L. C. Sinclair, E. Baumann, C. Cromer, G. Petron, C. Sweeney, P. P. Tans, I. Coddington, and N. R. Newbury, "Frequency-comb-based remote sensing of greenhouse gases over kilometer air paths," Optica 1, 290-298 (2014).
11. K. C. Cossel, E. M. Waxman, F. R. Giorgetta, M Cermak, I. R. Coddington, D. Hesselius, S. Ruben, W. C. Swann, G.-W. Truong, G. B. Rieker, and N R. Newbury, "Open-path dual-comb spectroscopy to an airborne retroreflector," Optica **4**, 724-728 (2017).
12. G. Ycas, F. R. Giorgetta, K. C. Cossel, E. M. Waxman, E. Baumann, N. R. Newbury, and I. Coddington, "Mid-infrared dual-comb spectroscopy of volatile organic compounds across long open-air paths," Optica 6, 165-168 (2019).
13. S. Coburn, C. B. Alden, R. Wright, K. Cossel, E. Baumann, G.-W. Truong, F. Giorgetta, C. Sweeney, N. R. Newbury, K. Prasad, I. Coddington, and G. B. Rieker, "Regional trace-gas source attribution using a field-deployed dual frequency comb spectrometer," Optica **5**, 320-327 (2018)
14. R. Grilli, G. Méjean, S. Kassi, I. Ventrillard, C. Abd-Alrahman, and D. Romanini, "Frequency Comb Based Spectrometer for in Situ and Real Time Measurements of IO, BrO, NO$_2$, and H$_2$CO at pptv and ppqv Levels," Environ. Sci. Technol. **46**, 10704-10710 (2012).
15. R. Grilli, M. Legrand, A. Kukui, G. Méjean, S. Preunkert, and D. Romanini, "First investigations of IO, BrO, and NO$_2$ summer atmospheric levels at a coastal East Antarctic site using mode-locked cavity enhanced absorption spectroscopy," Geophys. Res. Lett. **40**, 791-796 (2013).
16. B. Bernhardt, A. Ozawa, P. Jacquet, M. Jacquey, Y. Kobayashi, T. Udem, R. Holzwarth, G. Guelachvili, T. W. Hansch, and N. Picque "Cavity-enhanced dual-comb spectroscopy," Nature Photon. **4**, 55–57 (2010).
17. N. Hoghooghi, R. J. Wright, A. S. Makowiecki, W. C. Swann, E. M. Waxman, I. Coddington, and G. B. Rieker, "Broadband coherent cavity-enhanced dual-comb spectroscopy," Optica **6**, 28-33 (2019).
18. W. Zhang, X. Chen, X. Wu, Y. Li, and H. Wei, "Adaptive cavity-enhanced dual-comb spectroscopy," Photon. Res. **7**, 883-889 (2019).
19. L. Rutkowski and J. Morville, "Broadband cavity-enhanced molecular spectra from Vernier filtering of a complete frequency comb," Opt. Lett. **39**, 6664-6667 (2014).
20. L. Rutkowski and J. Morville, "Continuous Vernier filtering of an optical frequency comb for broadband cavity-enhanced molecular spectroscopy," J. Quant. Spectrosc. Radiat. Transf. **187**, 204–214 (2017).
21. A. T. Waz, G. Dudzik, P. R. Kaczmarek, and K. M. Abramski, "Multichannel WDM vibrometry at 1550 nm," Photonics Lett. Pol. **6**, 133-135 (2014).
22. A. Khodabakhsh, L. Rutkowski, J. Morville, and A. Foltynowicz, "Mid-infrared continuous-filtering Vernier spectroscopy using a doubly resonant optical parametric oscillator," Appl. Phys. B Lasers Opt. **123**, 210 (2017).
23. C. Lu, F. S. Vieira, F. M. Schmidt, and A. Foltynowicz, "Time-resolved continuous-filtering Vernier spectroscopy of H$_2$O and OH radical in a flame," Opt. Express **27**, 29521–29533 (2019).
24. I. E. Gordon et al., "The HITRAN2016 molecular spectroscopic database," J. Quant. Spectrosc. Radiat. Transf. **203**, 3-69 (2017).